\documentclass[
reprint,
nofootinbib,
amsmath,amssymb,amsbsy,
prd,
twocolumn,
superscriptaddress,
longbibliography,
preprintnumbers]{revtex4-2}
 
\usepackage{mathtools}

\usepackage[T1]{fontenc}
\usepackage{gfsartemisia-euler}
\usepackage{enumitem}
\usepackage{pifont}
\usepackage{amsmath}

\usepackage[mathscr]{euscript}

\makeatletter
\newcommand{\subalign}[1]{%
  \vcenter{%
    \Let@ \restore@math@cr \default@tag
    \baselineskip\fontdimen10 \scriptfont\tw@
    \advance\baselineskip\fontdimen12 \scriptfont\tw@
    \lineskip\thr@@\fontdimen8 \scriptfont\thr@@
    \lineskiplimit\lineskip
    \ialign{\hfil$\m@th\scriptstyle##$&$\m@th\scriptstyle{}##$\hfil\crcr
      #1\crcr
    }%
  }%
}
\makeatother

 \usepackage{lmodern}
\usepackage{graphicx}
\usepackage{hyperref}
\usepackage{amsfonts}
\usepackage[named]{xcolor}
\usepackage{tocvsec2}
\usepackage{slashed}
\usepackage{cancel}
\usepackage{comment}

\newcommand\beq{\begin{eqnarray}}
\newcommand\eeq{\end{eqnarray}}

\newcommand{\CD}{{\cal D}}

\newcommand\vev[1]{\langle #1 \rangle}

\newcommand{\mybar}[1]{\kern 0.6pt\overline{\kern -0.6pt#1\kern -0.6pt}\kern 0.6pt}

\def\U\Omega{U(1)_{\Omega}}

\hypersetup{
  colorlinks=true,   
  linkcolor=cyan,    
  citecolor=blue,    
  filecolor=magenta, 
  urlcolor=blue      
}
\usepackage[section]{placeins}

\begin{document}

\title{Weyl Fermions on a Finite Lattice}

\author{David B. Kaplan}
\email{dbkaplan@uw.edu}
\affiliation{Institute for Nuclear Theory, Box 351550, Seattle, Washington 98195-1550}

\author{Srimoyee Sen,}
\email{srimoyee08@gmail.com}
\affiliation{Department of Physics and Astronomy,  Iowa State University, Ames, Iowa 50011}

\date{\today}
\preprint{INT-PUB-23-046}

\begin{abstract}
The phenomenon of unpaired  Weyl fermions appearing on the sole $2n$-dimensional boundary of a $(2n+1)$-dimensional  manifold with massive Dirac fermions  was recently analyzed in Ref.~\cite{Kaplan2023z}. In this Letter we show that similar unpaired Weyl edge states can be seen on a finite lattice.  In particular, we consider the discretized Hamiltonian for a Wilson fermion in (2+1) dimensions with a 1+1 dimensional boundary and continuous time.  We demonstrate that the low lying boundary spectrum is indeed Weyl-like: it has a linear dispersion relation and definite chirality and circulates in only one direction around the boundary. We comment on how our results are consistent with Nielsen-Ninomiya theorem. This work removes one potential obstacle facing the program outlined in \cite{Kaplan2023z} for regulating chiral gauge theories.
\end{abstract}
 
\date{\today}

\maketitle

Understanding how to regulate chiral gauge theories on the lattice has been a longstanding challenge, brought into focus by the early work on anomalies by Karsten and Smit \cite{karsten1981lattice} and the no-go theorem of Nielsen and Ninomiya \cite{Nielsen:1980rz}.   The decades of failed attempts might have been enough to stop research in the field if it were not for the fact that the world is well described by a chiral gauge theory (the standard Model) and we need to fully understand   how it works nonperturbatively and how it can  defined. Furthermore, strongly coupled chiral gauge theories are expected to display  interesting phenomena  such as massless composite fermions and would be interesting to simulate numerically.  Recently it was proposed that Weyl fermions could appear as edge states on a finite five-dimensional  manifold with a single boundary, and that  a chiral gauge theory could be constructed on the four-dimensional boundary by gauging the exact vectorlike symmetry in the five-dimensional world.  In particular, the Dirac equation on a solid torus in odd dimensions was analyzed in the continuum and shown to support massless 2D or 4D Weyl fermions on the boundary. The problem was reduced to considering the Dirac equation on the 2D disk that forms the cross section of the torus, with the   coordinate around the edge of the disk  to be identified  with one of our spatial dimensions  \cite{Kaplan2023z} (for related work, see Ref.~\cite{aoki2023curved}). The proposal \cite{Kaplan2023z} exploits the fact that massless fermion edge states appear  at the boundary between two different topological phases, which can exist in discretized theories as well as the continuum, and as such are not subject to fine-tuning \cite{Kaplan:1992bt,Jansen:1992tw,Golterman:1992ub}.  Furthermore, it is clear that there exist nonlocal interactions in such a theory with a coefficient proportional to the gauge anomaly, so  it follows algebraically  that an anomalous  local 4D chiral gauge theory could never be constructed using this method, an important check on whether the method is sensible.  There  are several speculative components to the idea, and here we address one of the simplest: whether the free fermion spectrum discussed in the continuum in Ref.~\cite{Kaplan2023z} can be realized on a lattice.  The answer is yes, and we demonstrate that Weyl edge state modes are ubiquitous in extremely simple lattice theories of free fermions.

\begin{figure}[t]
    \includegraphics[width=.8\linewidth]{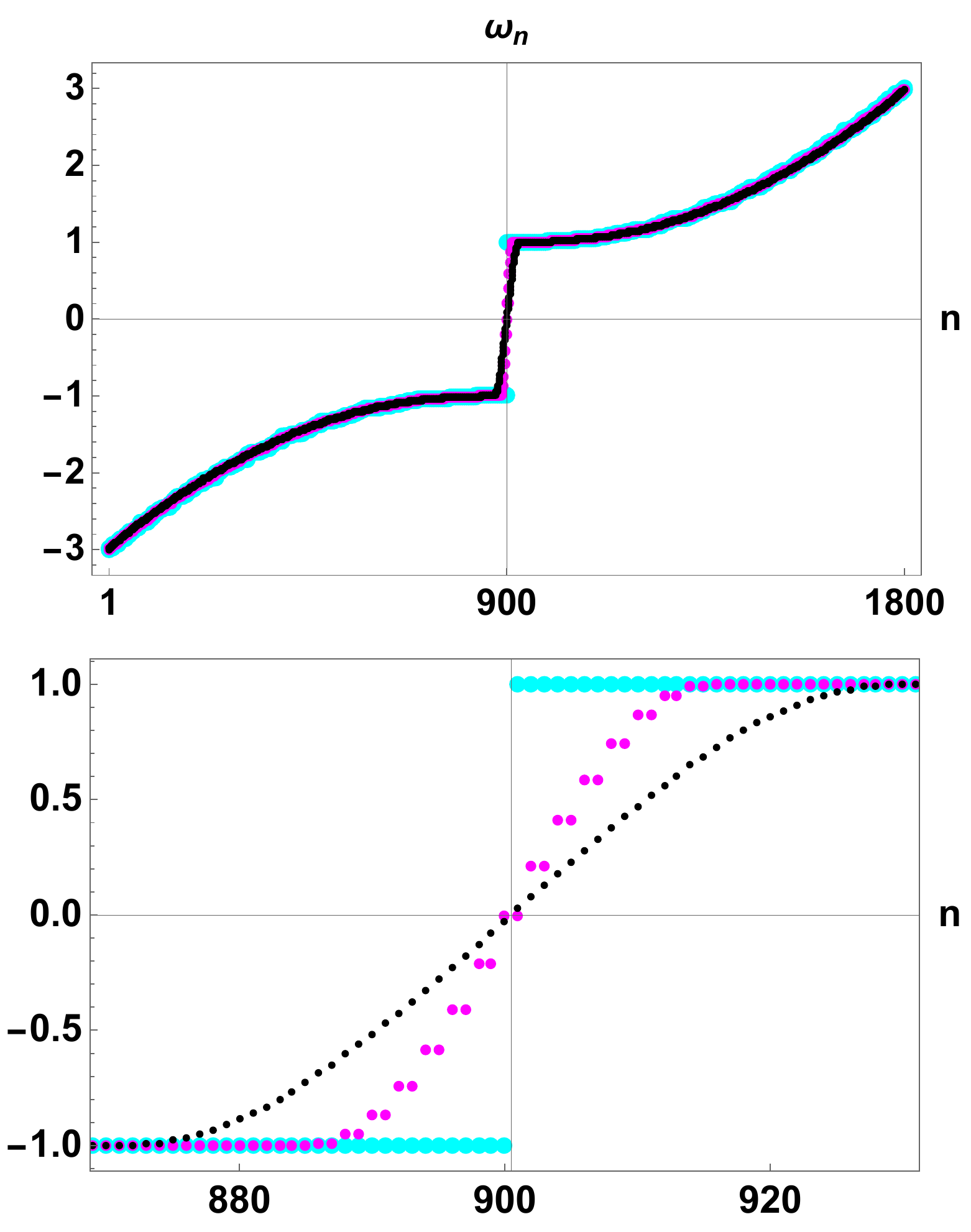}
    \caption{The 1800  ordered eigenvalues  $\omega_n$  on the $y$ axis  versus $n$ on the $x$ axis  for the free Wilson fermion Hamiltonian with $a=M=r=1$ on a $30\times 30$ lattice, with a magnified view of the crossing point in the  lower  panel. Cyan points are for purely periodic boundary conditions, corresponding to a spatial manifold with no boundary, and exhibit a gap. Magenta indicates mixed periodic and open boundary conditions, as specified by Shamir \cite{shamir1993chiral} for domain wall fermions, supporting a Weyl fermion on each of its two $S^1$ surfaces with opposite chiralities, exact zero modes, and evident degeneracy.  The nondegenerate black points are for purely open boundary conditions, representing the situation described in Ref.~\cite{Kaplan2023z}: a manifold with a single boundary (a square, in this case) supporting a single Weyl fermion without an exact zero mode.}
    \label{fig:comps}
\end{figure}


It is not obvious  that the continuum analysis should apply to the lattice. As a technical issue,  the chiral nature of the edge states in the continuum follows from discarding solutions to the Dirac eigenvalue equation which are divergent at the center of the disk.  If one considered an annulus with a hole in the center instead of a disk, in place of a single Weyl fermion edge state in the spectrum with level spacing proportional to the inverse disk radius, $1/R$, there would have been two of opposite chirality, the second solution circling the interior boundary of the annulus with level splitting set by the inner radius, $1/R'$. Since a lattice is full of holes, one should expect such mirror states when space is discretized, potentially ruining the mechanism. However, if we consider the lattice to be like the annulus, but with holes at the scale of the lattice spacing  $R'\sim a$, it may be reasonable to  expect  such mirror states  to be  gapped at the lattice cutoff. This gapping mechanism does not break the exact $U(1)$ symmetry [or $U(N_f)$ for multiple flavors], in which we hope to embed a chiral gauge group, as it does not result from a mass for the unwanted Weyl mode but instead from  forcing it to have high angular momentum around the small holes.   
 
This picture is simplistic but helps to address which of the four underlying assumptions   the Nielsen-Ninomiya theorem is being violated\footnote{The assumptions being that in the infinite volume limit the fermion operator be continuous and periodic in momentum space; that it be free of doublers (multiple zeros); that it have the correct continuum limit at zero wave number; and that it anticommute with $\gamma_5$.  The first assumption is required for locality in position space.}.  As seen in the continuum example, the chiral symmetry of the edge states on the disk descends directly from the exact fermion number symmetry of the higher dimension theory, and this is true on the lattice as well.   Therefore the chiral symmetry assumption is not being violated, the way it is by Ginsparg-Wilson fermions.  With the starting point being Wilson fermions in $d+1$ dimensions there are no doublers, and the correct continuum limit can be achieved at zero wavenumber, satisfying two more of the Nielsen-Ninomiya assumptions.  What is being violated  apparently is the assumption that the fermion operator be periodic and analytic in momentum. That assumption forbids the fermion energy levels as a function of momentum to cross zero an odd number of times, ruling out a Weyl fermion, while breaking it makes     a single Weyl mode possible. This is plausible in the picture of the annulus: the continuum limit is the large-$R$ limit, keeping $R'$ fixed.  Thus the modes on the outer edge develop a continuous spectrum while those of opposite chirality on the inner edge do not.

We show here that in fact such a picture seems to be realized.  What we find  is that a very standard lattice theory of Wilson fermions can describe a left-handed fermion bound on the outer rim of lattice manifold, which develops a continuous spectrum  at low energy as the size of the lattice is taken to be large. There are no long wavelength states one might identify as a mirror Weyl fermion.

In this Letter we examine the spectrum for an extremely simple system: the  discretized Hamiltonian for free Wilson fermions in $2+1$ dimensions,
\beq
H &=& \gamma_0 \CD= \gamma_0\,\sum_{\mu=1}^2\gamma_\mu\partial_\mu  + M + \frac{r}{2} \Delta\ ,\cr
\gamma_0 &=& \sigma_3\ ,\quad \gamma_1 = \sigma_1\ ,\quad \gamma_2=\sigma_2
\ .
\label{e1}
\eeq
The   $\partial_\mu$ are symmetric lattice derivatives with $\mu = 1,2$, and $\Delta$ is  the 2D lattice Laplacian (see the Supplemental Material for details of our lattice implementation).
Since the features of the fermion spectrum we are interested in are topological in nature, they are determined by both the ratio $M/r$,  \cite{Kaplan:1992bt,Jansen:1992tw,Golterman:1992ub} and by the boundary conditions \cite{shamir1993chiral}.   We take $M=r=1$, ensuring a nontrivial topological phase in the bulk.  For the boundary conditions, as discussed in  Ref.~\cite{Kaplan2023z},  a $(d+1)$-dimensional manifold $(d=2n)$ with no boundaries is expected to support no massless  states; a $(d+1)$ manifold with two disconnected boundaries is expected to support a light Dirac fermion (two Weyl fermions of opposite chirality with exponentially small interactions), and a $(d+1)$ manifold with a single boundary should have a single massless Weyl fermion with exponentially small nonlocal interactions. These considerations are independent of details of the shape of the lattice or whether the low energy theory possesses $d=1+1$ Lorentz invariance.  Therefore we start by considering the simplest possible lattice, a square lattice, cut into a square shape of dimension $L\times L$, with three possible boundary conditions:  
(i)  periodic boundary conditions in both coordinates, $\psi(x+L,y) =\psi(x,y) $, $\psi(x,y+L) =\psi(x,y) $.  This is the usual boundary condition for Wilson fermions, the lattice has the topology of a 2-torus without boundary, and the spectrum is gapped without any continuum low lying modes.
(ii)  Periodic boundary conditions in one direction and open boundary conditions in the other:  $\psi(x+L,y) = \psi(x,y)$ and $\psi(x,0) = \psi(x,L+1) = 0$.  This is the prescription for conventional domain wall fermions proposed by Shamir \cite{shamir1993chiral}.  In this case the lattice has the topology of an open cylinder with a circle for the boundary at each end.  With two disconnected pieces to the boundary,  the lattice supports two Weyl fermions with opposite chirality.  They have an interaction vanishing exponentially with the length of the cylinder, turning them into a very light Dirac fermion with mass exponentially small in $ML$, and have no  exact chiral symmetry to be naively gauged\footnote{This system has been fully described in the literature.  In the large volume limit the massless Dirac mode is described by the overlap operator \cite{Neuberger:1997fp,neuberger1998vectorlike,Narayanan:1993sk,Narayanan:1993zzh,Narayanan:1993ss}. In turn, the overlap operator solves the Ginsparg-Wilson equation \cite{Ginsparg:1981bj} which clarifies exactly how the mode violates chiral symmetry, even in the infinite volume limit, which can be thought of as being due to the nondecoupling of massive modes in the bulk which generate a Chern-Simons form, which accounts for the anomaly \cite{Callan:1984sa}.}.
(iii) Open boundary conditions in both directions, $\psi(0,y)=\psi(L+1,y) = \psi(x,0) = \psi(x,L+1)=0$.  This is the case of interest, realizing a single massless Weyl mode in its spectrum.\footnote{Open boundary conditions have previously been considered in a different context in Ref.~\cite{luscher2013lattice}}   
The expected behavior is confirmed by plotting the eigenvalues of $H$ for these three different boundary conditions, as shown in Fig.~\ref{fig:comps}.

 For a realistic simulation of a $1+1$-dimensional  theory of a Weyl fermion, one would want to equate spatial translations in the continuum theory with rotations of the two-dimensional lattice Hamiltonian.  Even though short distance modes will always be sensitive to the symmetry of the underlying square lattice, to study the appropriate long wavelength edge modes  one will need to consider a square lattice cut in the shape of a disk. 
 To construct the lattice Hamiltonian on such a disk we start with an $L\times L$ square lattice with open boundary conditions, and then  act on both sides of the Hamiltonian with projectors setting to zero the field on all sites at radius $r\ge L/2-1$, discarding eigenvectors corresponding to those sites.  Alternatively, one can use the square lattice with a radially dependent mass with $M=1$ for $r< L/2$ and $M\ll -1$ for $r\ge L/2$, discarding eigenvectors with $\vert\omega_n\vert \gg 1$.  Both methods give similar results.  As shown in Ref.~\cite{Kaplan2023z}, linear momentum of the edge state is then given by $J/R$, where $J$ is the angular momentum operator  $J = (\vec r\times\vec p + S)$, where $S = i/4[\gamma_x,\gamma_y]$.   


\begin{figure}[t]
    \centering
    \includegraphics[width=1.0\linewidth]{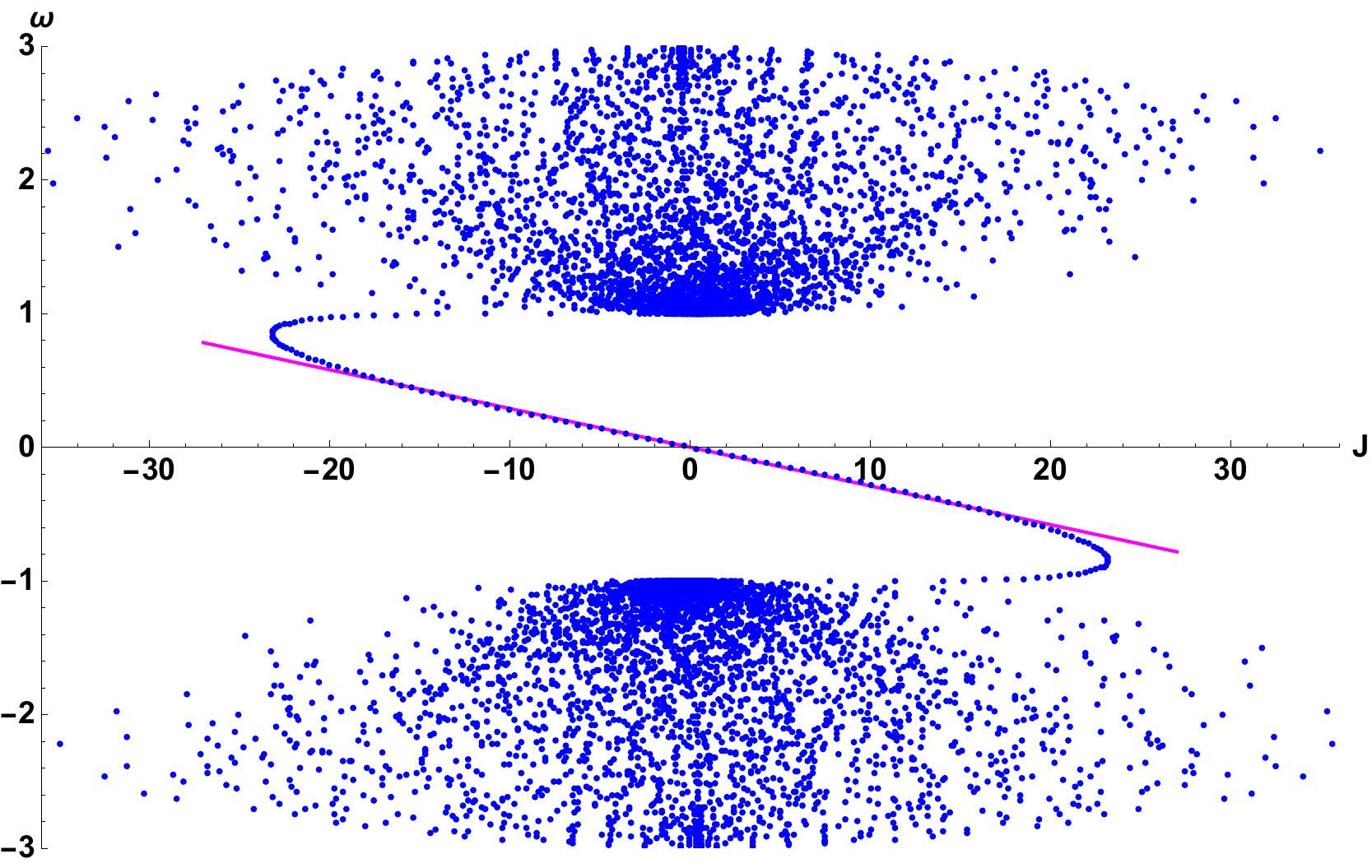}
    \caption{The energy for every mode plotted versus the expectation value of the angular momentum $J$   on a  radius $R= 34$ disk of square lattice with open boundary conditions.  The  straight line is a plot of the function $\omega = - \vev{J}/R$, where $\vev{J}/R$ can be identified as the edge state momentum along the boundary. }
    \label{fig:jplot}
\end{figure}

 \begin{figure}[t]
     \centering
     \includegraphics[width=.9\linewidth]{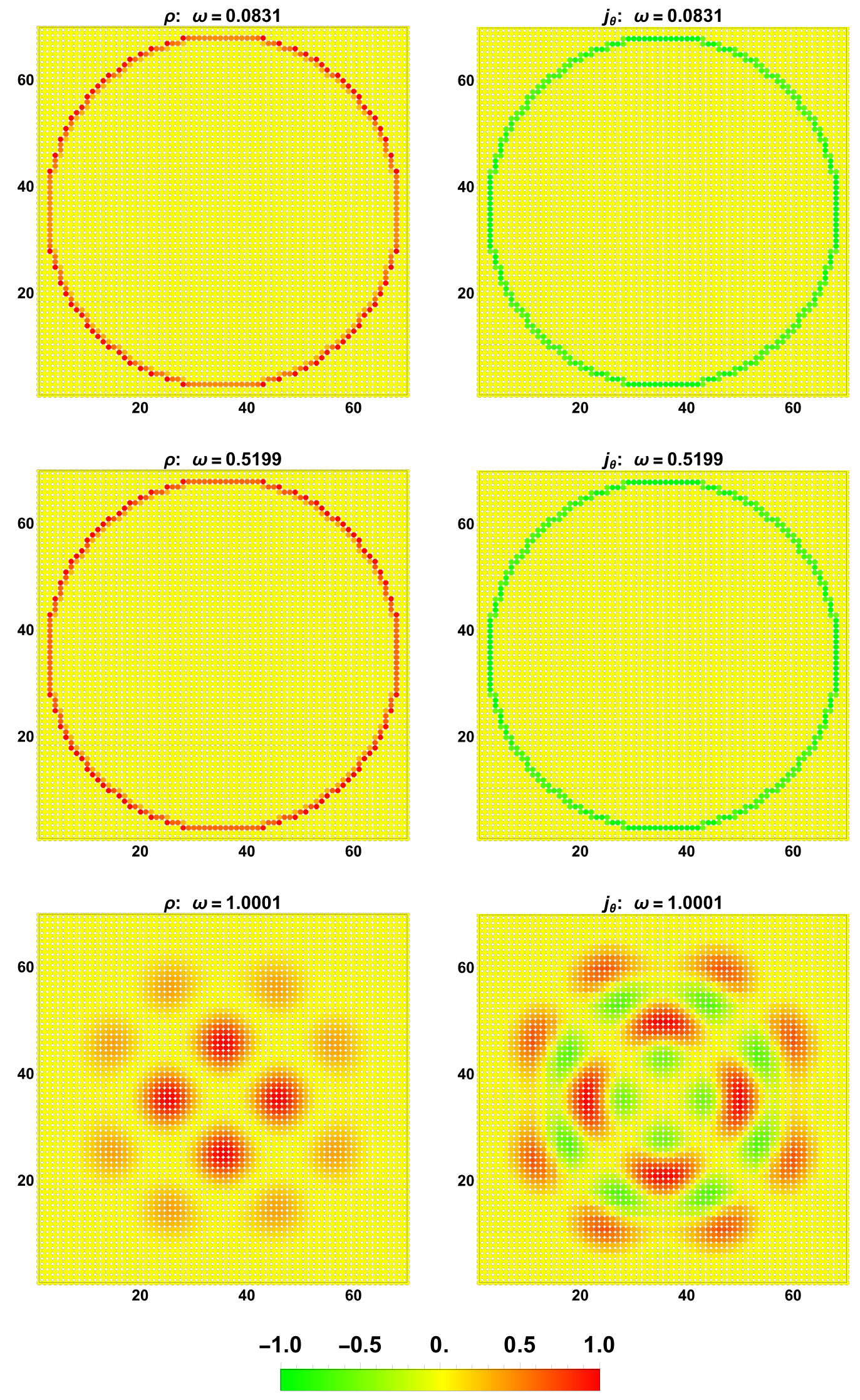}
     \caption{The fermion number charge densities $\rho$ (left) and angular fermion number current density $j_\theta$  (right) on  a  radius $R= 34$ disk of square lattice with open boundary conditions for energy eigenstates with energies $\omega = 0.0831, 0.5199, 1.0001$ in lattice units (top to bottom). Both the $\rho$ and $j_\theta$ plots display sharp edge localization for the two modes in the Weyl gap $\vert\omega\vert<1$, and  delocalization for the mode slightly above $\omega=1$. The Weyl modes exhibit strictly negative  $j_\theta$ corresponding to clockwise rotation around the lattice boundary, while the bulk mode does not display any net chirality. The data has been rescaled by the maximum magnitude of the density value. }
     \label{fig:density}
 \end{figure}

There are several properties we expect to find  belong to the Weyl modes in the gap $-1\le \omega_n\le 1$. (i)   Their energy eigenvalues should obey the dispersion relation appropriate for a massless chiral fermion, $\omega= -\vev{J}/R$ (with $O(1/MR)$ corrections), where $\vev{J}$ is the total angular momentum quantum number. (ii) These Weyl states should be localized on the boundary. (iii) The edge states should be chiral, only rotating clockwise around the boundary in the simple model we consider.  States of opposite chirality can be obtained by reversing the signs of both $M$ and $r$ in $H$.  

The expected dispersion relation characteristic of a massless Weyl fermion is evident in Fig.~\ref{fig:jplot}, where we have plotted the energy of each energy eigenstate versus the expectation value of the total angular momentum operator $J$ for a disk-shaped  lattice with radius $R=34$ sites. (For this calculation we used improved derivative operators for  the orbital part of $J$, canceling lattice artifacts to $O(a^7)$, thereby gaining a modest improvement in the range of linear behavior in $\omega$ versus $-\vev{J}/R$.) The Weyl modes have an increasingly dense spectrum as one increase $R$, but rather then being periodic in $J$, the spectrum turns back at large energy and dissolves into a sea of lattice artifact states. 
This result  contradicts  the conventional wisdom derived from the Nielsen-Ninomiya theorem that a Weyl-like dispersion relation    cannot arise from a physically sensible microscopic lattice theory of free fermions.

In Fig.~\ref{fig:density}  we show the density plots  for three positive energy eigenstates of  $j_0$ and $j_\theta$, where $j_\mu$ is the lattice construction of the continuum fermion number current operator $\bar\psi\gamma_\mu\psi$, and $j_\theta = (-\sin\theta j_1+ \cos\theta j_2)$ -- one with $\omega\ll 1$, one in the middle of  the $\vert\omega\vert <1$ ``Weyl gap'', and one just above the edge of the gap.  The plots confirm expectations: the first two modes are  well localized at the boundary and exhibit counterclockwise current flow, the latter very much a bulk state with no definite chirality.

We have  contrasted the three simplest possible boundary conditions on a two-dimensional square lattice of Wilson fermions  in one of their nontrivial topological phases: (\{periodic, periodic\}, \{periodic, open\}, and \{open, open\}). We have demonstrated that they lead to quite different physics: a gapped spectrum of no interest for a continuum theory in the \{periodic, periodic\} case, the well studied case of conventional domain wall fermions with an almost massless Dirac mode on the surface in the  \{periodic, open\} case, and a new example with a massless Weyl fermion on the surface for \{open, open\}. When the latter is  reformulated as a lattice cut in the approximate shape of a disk with an open boundary condition, the Weyl states display approximate rotational invariance as required if one hopes to attain Lorentz invariance in the IR. Generalizations to higher dimension lattices are expected to lead to even richer phase structure.  It is gratifying that insights from the continuum study of a manifold with a single boundary supporting a Weyl fermion carry over directly to the lattice, and that the theory has no problem sidestepping restrictions implied by the Nielsen-Ninomiya theorem.  There is still much that can be learned about the lattice discussed here, such as  the degree of nonlocality in the propagation of the Weyl surface mode, and whether there is some version of  the overlap operator to describe its behavior without reference to the higher dimension world. Ultimately, a realistic computation will involve discretizing time as well as space and including  gauge fields as  link variables, integrating over their boundary values as discussed in  Ref.~\cite{Kaplan2023z}, the latter entailing additional questions and challenges not addressed here.

\bigskip
\centerline{\bf Acknowledgements}
\bigskip
We wish to thank M. Golterman, Y. Shamir, and S. Sharpe for helpful comments. D.B.K. is supported in part by DOE Grant No. DE-FG02-00ER41132. S.S.~acknowledges support from the U.S. Department of Energy,
Nuclear Physics Quantum Horizons program through the
Early Career Award DE-SC0021892.

\bigskip

\bigskip
\centerline{\bf APPENDIX}
\bigskip

We provide here a brief account of those details of the lattice calculations employed in our paper which appear as Supplemental Materials in the published version. 

We use the term ``square lattice'' to refer to a conventional lattice whose fundamental cell is a square; this is the only sort of lattice considered in this paper.  By a lattice ``cut into a square'' or ``cut into a disk '' we are describing the boundaries of the lattice.  In the former case we describe a  lattice with $L$ sites in each direction, with various possible boundary conditions. In the latter case what we did operationally was first define a projection operator $P_R$ with the property
\beq
P_R\psi(x) = \begin{cases}0 & x\ge R \cr \psi(x) & x< R\ ,
\end{cases}
\eeq
and then we defined the Hamiltonian on the disk to be
\beq
H_\text{disk } = P_R H_{L\times L} P_R\ ,
\eeq
where $H_{L\times L} $ is the Wilson fermion Hamiltonian on an $L\times L$ square lattice and $R<L/2$.   We  then computed the eigensystem for $H_\text{disk } $, and confirmed that all eigenvectors with  exactly  zero eigenvalues corresponded to states outside the disk.  We discarded the corresponding eigenvectors, retaining those with nonzero eigenvalue to span our Hilbert space. For
Figs.~\ref{fig:jplot}, \ref{fig:density} we used $L=70$ and $R=34$.  The resulting lattice is shown in Fig.~\ref{fig:disclattice}.

\begin{figure}[b]
\setcounter{figure}{3}
    \centering
    \includegraphics[width=.4\linewidth]{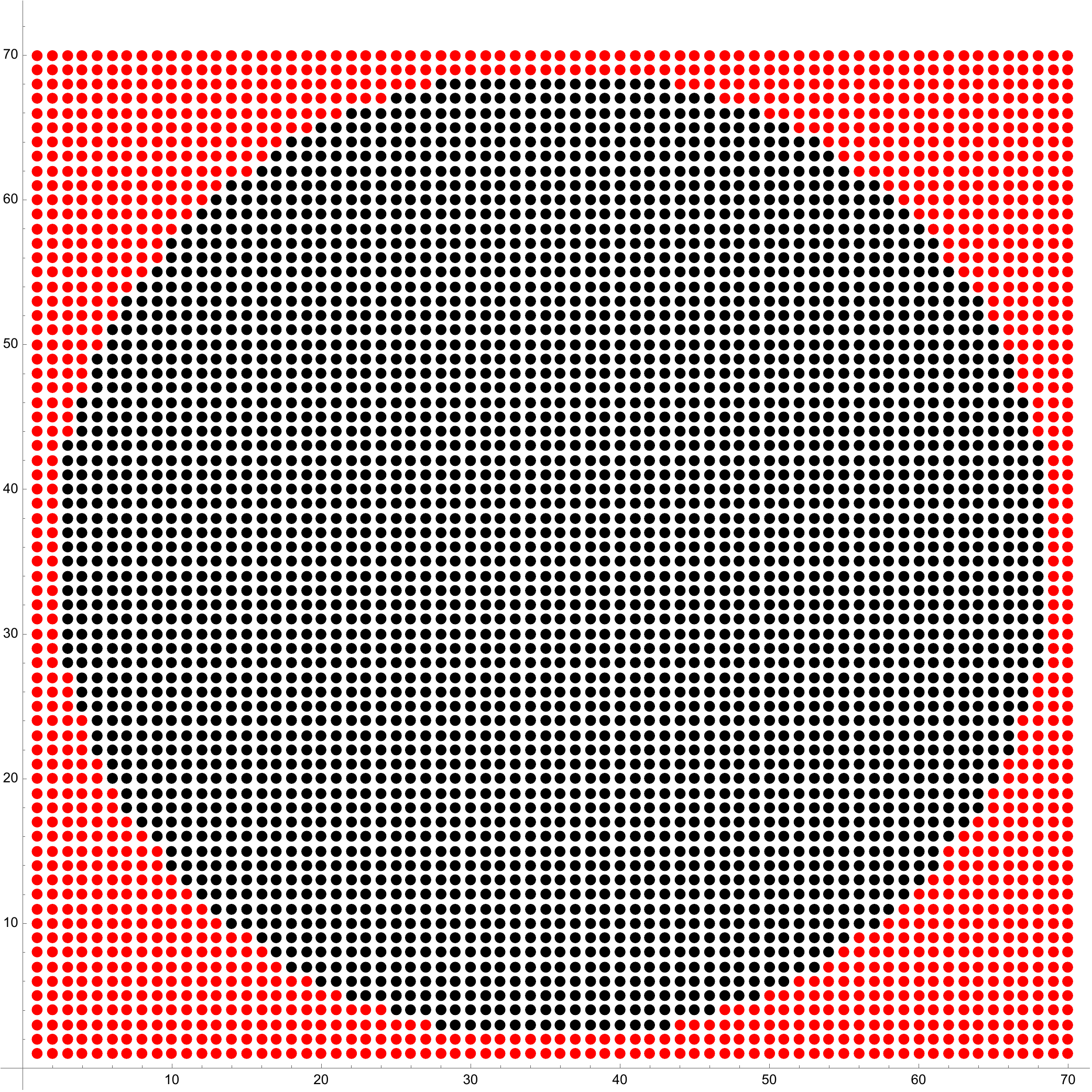}
    \caption{The lattice cut into an approximate disk used for the calculation of Figures ~\ref{fig:jplot}, \ref{fig:density} with $L=70$, $R=34$. The red points are those on the $L\times L$ lattice that were excluded, while the black ones were kept.    }
    \label{fig:disclattice}
\end{figure}

The derivatives in eq. \ref{e1} are defined to be:
\beq
\partial_\mu \psi(x) &=&  \frac{\psi(x + a \hat\mu) -  \psi(x - a \hat\mu)}{2a}\ ,\cr&&\cr
\Delta  \psi(x) &=& \sum_{\mu}\frac{ \psi(x + a \hat\mu) -2 \psi(x) + \psi(x - a \hat\mu)}{a^2}\ ,
\eeq
where $\hat \mu$ is a unit vector in the $\mu$ direction.  Fig.~\ref{fig:comps} resulted
from computing the eigenvalues of $H$ on an $L\times L$ square lattice with the boundary conditions described in the text.

As described there, the calculations for Figs.~\ref{fig:jplot}, \ref{fig:density} of the text were performed using $H_\text{disk }$.  For Fig.~\ref{fig:jplot}, for each eigenvector of $H_\text{disk }$ we plotted the energy eigenvalue versus the expectation value of the angular momentum operator $J$, given in the text of the paper as
\beq
J = \vec r \times \vec p + \frac{i}{4}\left[\gamma_x,\gamma_y\right]\ ,
\eeq
where $\gamma_{x,y}$ were the Pauli matrices in eq.~\ref{e1} of the paper.  For $\vec r$ we took the coordinate on the lattice, while for $\vec p$ we took $p = -i\vec\nabla$, where $\vec \nabla$ is a lattice derivative.  As mentioned in the text, we found the Weyl branch of the plot was a better fit to a straight line over a marginally improved range of energy  if we used an improved lattice derivative, and for Fig.~\ref{fig:jplot} we used
\beq
\nabla_\mu \psi(x)&=&\frac{1}{a}\left[\frac{75}{128} \psi(x+a\hat\mu)-\frac{25}{768} \psi(x+3a\hat\mu)\right.\nonumber\\
&&\left.+ \frac{3}{1280}\psi(x+5a\hat\mu) -( a\to -a)\right]
\eeq
which gives the continuum derivative up to $O(a^7)$ corrections. We do not pretend to be performing a systematic elimination of lattice artifacts up to $O(a^7)$.

For Fig.~\ref{fig:density}, for the density $\rho$ we plotted  $|\xi(x,y)|^2$, where $\xi$ is the eigenvector of $H_\text{disk }$ for the three energy eigenvalues indicated as $\omega_n$ in the figure. We then rescaled the function by its maximum value on the lattice before plotting, so that it ranged on the interval $[0,1]$.  The positive value is appropriate for the positive energy eigenstates we examined. For the current $j_\theta$ we defined
\beq
j_\mu(x)&=&-\text{Im}\left[ \xi^\dagger(x) \gamma_0 (1+\gamma_\mu)\xi(x+a\hat\mu)\right]\nonumber\\
&&-\text{Im}\left[\xi^\dagger(x-a\hat\mu) \gamma_0 (1+\gamma_\mu)\xi(x)\right]\ .
\eeq
 The terms in this expression correspond to what a vector gauge boson would couple to in $H_\text{disk }$, averaged over the links on either side of the site $(x,y)$.  We then computed the angular current as
 \beq
 j_\theta(x,y) = \frac{-\bar y j_x +\bar x j_y}{\sqrt{\bar x^2+\bar  y^2}}\ ,
 \eeq
where $\bar x$ and $\bar y$ are the lattice coordinates $\{x,y\}$ shifted by $\left\{\frac{L+1}{2},\frac{L+1}{2}\right\}$ so that they vanish at the center of the disk.  We then rescaled $j_\theta$ by its maximum absolute value, so that it ranged on the interval $[-1,1]$. 
 
\bibliography{refs.bib}
\end{document}